\documentstyle[12pt]{article}
\begin{document}
\title{\bf Spectral Geometry of Operator Polynomials and Applications to QFT}

\bigskip\bigskip
\bigskip\bigskip
\bigskip\bigskip
\bigskip\bigskip

\author{       Dmitri V. Fursaev ~\thanks{
        fursaev@thsun1.jinr.ru}  \\ {}\\
     {\small\it Bogoliubov Laboratory of Theoretical Physics}\\
{\small \it  Joint Institute for Nuclear Research}\\
{\small \it  141980 Dubna, Moscow Region, Russia}
\\
}
\maketitle
%%%%%%%%%%%%%%%%%%%%%%%%%%%%%%%%%%%%%%%%%%%%%%%%%%%%%%%
\begin{abstract}
A class of non-linear eigenvalue problems defined in the form
of operator polynomials is investigated. The problems are related
to wave equations
which appear in a relativistic quantum field theory.
Spectral asymptotics for this class are found explicitly.
The properties of
operator polynomials are analyzed for scalar, spinor and gauge
fields.
It is also shown how to use these results in finite temperature theories.
\end{abstract}
%%%%%%%%%%%%%%%%%%%%%%%%%%%%%%%%%%%%%%%%%%%%%%%%%%

\bigskip
%\vspace{3cm}

\bigskip

\baselineskip=.6cm


\newpage

\section{Introduction}
\setcounter{equation}0

Consider a class of eigenvalue problems
which have the following form:
\begin{equation}
\label{1}
P(\omega)\phi_\omega\equiv (P_n+\omega P_{n-1}+...+\omega^{n-1}P_1 +
\omega^n P_0)\phi_\omega=0~~,
\end{equation}
where $\omega$ is a complex spectral parameter and $\phi_\omega$ is an eigenfunction.
$P_k$ ($k=0,..., n$)
are partial differential operators of $k$-th order
acting on a Hilbert space ${\cal H}$.
$P(\omega)$ represents an example of what is called
the operator polynomial or polynomial
operator pencil \cite{Markus}. In general
$P_k$ and $P_l$ do not commute if $k\neq l$ and (\ref{1})
is essentially non-linear problem with respect to the eigenvalue
$\omega$.

Problems like (\ref{1}) appear in different applications (oscillations in a viscous fluid,
Schr\"odinger equation with energy dependent potential and etc), see \cite{Markus}
for the introduction and references. The focus of the present work
is on applications related to quantum field theory.

Consider equation of motion for a free field $\phi(x,x^0)$ ($x$ and $x^0$ denote
spatial and time coordinates, respectively). It can be always
written as
\begin{equation}
\label{2}
P(i\partial_0)\phi(x,x^0)\equiv (P_2+P_{1}(i\partial_0) +
P_0(i\partial_0)^2)\phi(x,x^0)=0~~,
\end{equation}
where $\partial_0=\partial /\partial_0$. Operators $P_k$ depend only on spatial derivatives
and may have additional indexes related to a spin-tensor structure.
Field $\phi$ propagates in an external classical background.
If the background is stationary the variables can be separated by choosing
$\phi(x,x^0)=e^{-i\omega x^0}\phi_\omega(x)$. From (\ref{2}) one gets
\begin{equation}
\label{3}
P(\omega)\phi_\omega(x)= (P_2+ \omega P_{1} +\omega^2 P_0)\phi_\omega(x)=0~~.
\end{equation}
This is a quadratic operator polynomial, a particular case of problem (\ref{1}).

Eigenvalues $\omega$ determine the spectrum of energies of single-particle
excitations which is used to
compute a number of important physical quantities, like the vacuum (Casimir) energy
$E=\pm \frac 12 \sum |\omega|$, free energy and etc.

Consider for instance a charged Dirac field $\psi$  in an external static
electromagnetic field with the potential $A_\mu=A_\mu(x)$. The equation is
\begin{equation}\label{4}
[\gamma^\mu D_\mu+m]\psi=0~~,
\end{equation}
where $D_\mu=\nabla_\mu -ie A_\mu$ and $e$ is the charge. To get  for the
single-particle spectrum problem like (\ref{3})
one has to act by the operator $[\gamma^\mu D_\mu-m]$
on the left hand side (l.h.s.) of (\ref{4}) and make substitution
$\psi(x,x^0)=e^{-i\omega x^0}\psi_\omega(x)$. This yields
\begin{equation}\label{5}
[D_k D^k-m^2+e^2A_0^2-
{e \over 2} ({\cal F}-2i\gamma^0\gamma^k\partial_k A_0)+
\omega~ 2e A_0 + \omega^2 ]\psi_\omega(x)=0~~,
\end{equation}
where ${\cal F}=(\partial_jA_k-\partial_k A_j)\gamma^j\gamma^k$ and $k,j=1,2,3.$
When the time component $A_0$ of the potential  is a function
of spatial coordinates (\ref{5}) cannot be reduced to
a standard eigenvalue problem because operator $P_1=2e A_0$ does not commute with $P_2$.

Sec. 2 of this contribution contains a summary of  results
obtained in \cite{df:01,df:02a}. The aim is to show that the notion of
spectral geometry can be extended to a class of quadratic operator
polynomials which appear in problems of quantum field theory.
Applications to finite temperature theories are discussed in Sec. 3 where, as
an example, the obtained results are
used to describe the Debye screening in hot plasma.
Finally, some important features of operator polynomials for scalar, spinor and gauge
fields are analyzed in Sec. 4.

\section{Spectral Geometry of Quadratic Polynomials}
\setcounter{equation}0

To describe the results we rewrite (\ref{3}) in the form
\begin{equation}\label{6}
\left[\omega^2-L(\omega)\right]\phi_\omega(x)=0~~,
\end{equation}
\begin{equation}\label{7}
L(\omega)=L_2+\omega L_1+\omega^2 L_0~~,
\end{equation}
where $L(\omega)=\omega^2-P(\omega)$ and
$L_k$ are partial differential operators of the $k$-th order.
Suppose for a moment that $\omega$ is a complex
parameter and consider a one-parameter family of operators $L(\omega)$.
Let $\Lambda_k(\omega)$ be eigenvalues of $L(\omega)$ for a given $\omega$,
\begin{equation}\label{8}
L(\omega)\phi^{(\omega)}_{\Lambda_k}=\Lambda_k(\omega)\phi^{(\omega)}_{\Lambda_k}.
\end{equation}
If the spectrum $\Lambda_k(\omega)$ is known for any $\omega$,
the spectrum of problem (\ref{6}) is determined by roots of equation
\begin{equation}\label{9}
\chi(\omega,\Lambda_k)=0,
\end{equation}
\begin{equation}\label{10}
\chi(\omega,\Lambda_k)=\omega^2-\Lambda_k(\omega).
\end{equation}
Denote these roots by $\omega_{k,i}$
and define the following function on the spectrum
\begin{equation}\label{11}
\chi'(\omega_{k,i})=\partial_\omega\chi(\omega_{k,i},\Lambda_k)~~\mbox{where}~~~
\omega_{k,i}^2=\Lambda_k(\omega_{k,i}).
\end{equation}
It is assumed that at first the derivative in $\chi'(\omega_{k,i})$
over $\omega$ is taken for the fixed
branch of eigenvalues $\Lambda_k(\omega)$ and after that the result
is considered on the corresponding root $\omega_{k,i}$.
In what follows we write  $\omega$ instead of
$\omega_{k,i}$.

We  make three assumptions:
i) $L(\omega)$ is a Laplace type operator of the form:
\begin{equation}\label{7a}
L(\omega)=-(\nabla_k+iA_k+i\omega a_k)(\nabla^k+iA^k+i\omega a^k)
+\omega B+V
\end{equation}
which acts on sections
to vector bundles over a $d$-dimensional compact Riemannian manifold
${\cal M}_d$; ii) the spectrum of $L(0)$ is strictly positive;
iii) the function $\chi'(\omega)$, see (\ref{11}), is positive (negative)
on positive (negative) eigenvalues $\omega$.

First two conditions are technical (see details in \cite{df:02a}).
The origin of condition (iii) is explained in Sec. 4.
At this point we note that in the limit when the polynomial is trivial,
$L(\omega)=L_2$, this condition is
satisfied because $\chi'(\omega)=2\omega$.

To describe the spectrum of (\ref{6}) at large $|\omega|$ we introduce the
pseudo-trace
\begin{equation}\label{16}
K(t)=\frac 12 \sum_\omega e^{-t\omega^2},~~~t>0,
\end{equation}
where summation goes over the real eigenvalues $\omega$.
The asymptotics of $K(t)$ at small  $t$
is related to the distribution of large $|\omega|$.
Given conditions (i)--(iii) the following asymptotic series
takes place  at small  $t$ \cite{df:02a}
\begin{equation}\label{17}
K(t) \sim {1 \over (4\pi t)^{d/2}}\sum_{n=0}^\infty \left[a_{n} t^{n}+
b_{n} t^{n+1/2}\right].
\end{equation}
$a_n$ and $b_n$
can be computed by using expansion formula for the heat kernel
\begin{equation}\label{18}
K_\omega(t) =\mbox{Tr}~e^{-tL(\omega)}
\sim {1 \over (4\pi t)^{d/2}}\sum_{n=0}^\infty
\left[a_{n}(\omega)t^{n} +b_{n}(\omega)t^{n+1/2}\right]
\end{equation}
which is well-known \cite{Gilkey}. For operators defined in (\ref{7a}),
\begin{equation}\label{19}
a_{n}(\omega)=\sum_{m=0}^{n}a_{m,n}\omega^m,~~~
b_{n}(\omega)=\sum_{m=0}^{n}b_{m,n}\omega^m
\end{equation}
and it can be shown that \cite{df:02a}
\begin{equation}\label{20}
a_n=\sum_{m=n}^{2n}(-1)^{n-m}{\Gamma\left(-\frac d2+m\right)
\over \Gamma\left(-\frac d2+n\right) } a_{2(m-n),m},
\end{equation}
\begin{equation}\label{21}
b_n=\sum_{m=n}^{2n}(-1)^{n-m}{\Gamma\left(-{d-1 \over 2}+m\right)
\over \Gamma\left(-{d-1 \over 2}+n\right) } b_{2(m-n),m}.
\end{equation}
Coefficient $a_0$ is proportional to the volume of ${\cal M}^d$,
other $a_n$, $b_n$ are integrals of local functionals of background fields
and $b_n$ are non-trivial when ${\cal M}^d$ has a boundary.
The notion of spectral geometry does extend
to a class of quadratic operator polynomials.

The proof of (\ref{17}),  (\ref{20}), (\ref{21}) is based on relation \cite{df:02a}
\begin{equation}\label{46}
K(t)={1 \over 4\pi}
\int^{\infty}_{0}d\omega~
\int_{C}
dz ~e^{-\omega^2(t-iz)}
\left(2\omega+{1 \over iz}\partial_\omega\right)(K_\omega(iz)+K_{-\omega}(iz)),
\end{equation}
where a contour $C$ goes in the complex plane just below the real axis
$\mbox{Im}~z=0$. Thus $K(t)$ is determined by part of $K_\omega(t)$ which
is  symmetric with respect to the transformation $\omega$ to $-\omega$.
This property is a direct consequence of condition (iii) which we imposed on
the spectrum of the operator polynomial. This also explains why
coefficients $a_n$ and $b_n$ in (\ref{20}), (\ref{21}) are determined
by the symmetric part of coefficients $a_n(\omega)$ and $b_n(\omega)$.

\section{Applications to Finite-Temperature Theories}
\setcounter{equation}0

Spectral asymptotics of quadratic operator polynomials can be used in a number
of physical applications. One of them is studying the  behaviour of
the free energy of non-interacting quanta at high temperatures. The
free energy for Fermi particles at temperature $T=\beta^{-1}$ is defined as
\begin{equation}\label{22}
F(\beta)=- \beta^{-1} \sum_k\ln \left(1+ e^{-\beta E_k}\right).
\end{equation}
Summation goes over single-particle energies $E_k=|\omega_k|$ which are
determined by
the real eigenvalues $\omega_k$ of the corresponding wave equations.

It is convenient to start with the case
when the single-particle spectrum is determined by the standard
linear eigenvalue problem, i.e., when operator $L(\omega)=L_2$ and
it does not depend on $\omega$. This case is well studied.
It is known that in three dimensions
$d=3$
the free energy
in the high temperature limit has the following asymptotic expansion \cite{dk}
\begin{equation}\label{28}
F(\beta,A)\sim -{7\pi^2 \over 720} T^4 a_0-{1 \over 48} T^2 a_1-
{1 \over 16\pi^2}\ln (T/\varrho)a_2+O(T^{-2})~~
\end{equation}
(we neglect boundary terms for simplicity).
Here $a_n$ are the coefficients of the asymptotic expansion of
the trace $\mbox{Tr}~e^{-tL_2}$, $\varrho$ is a dimensional parameter.

In the high temperature limit $F(\beta)$ is
determined by behaviour of the single-particle spectrum at large
$|\omega|$. For polynomial (\ref{6})
this behaviour is described by asymptotic formula
(\ref{17}) which has exactly the same form as
expansion of
$\mbox{Tr}~e^{-tL_2}$. Therefore, for systems whose spectrum is determined by
(\ref{6}) formula (\ref{28}) can be used if $a_n$ are calculated
with the help of (\ref{20}) (analogously for boundary terms if they are
present).

Let us use this method to derive the free energy of electron-positron plasma
in an external static electromagnetic field $A_\mu(x)$.
(Finite-temperature theories in gravitational backgrounds are discussed
in \cite{df:01}).
Coefficients
$a_n$ are determined by (\ref{20}) for problem (\ref{5}). The
corresponding operator is
\begin{equation}\label{5a}
L(\omega)=
-{\cal D}_k{\cal D}^k+
m^2-e^2A_0^2+
{e \over 2} ({\cal F}-2i\gamma^0\gamma^k\partial_k A_0)-
2e A_0\omega~.
\end{equation}
By using standard results \cite{Gilkey} one easily finds
coefficients of $L(\omega)$ and obtains
\begin{equation}\label{30}
a_0=a_{0,0}=4\int_{{\cal M}^3} d^3x~~,
\end{equation}
\begin{equation}\label{31}
a_1=a_{0,1}+\frac 12 a_{2,2}=4\int_{{\cal M}^3} d^3x (-m^2+2e^2 A_0^2)~~.
\end{equation}
\begin{equation}\label{32}
a_2=a_{0,2}-\frac 12 a_{2,3}+\frac 34 a_{4,4}=
{2e^2 \over 3} \int_{{\cal M}^3} d^3x F_{\mu\nu}F^{\mu\nu}~,
\end{equation}
$F_{\mu\nu}=\partial_\mu A_\nu-\partial_\nu A_\mu$.
Consider the effective thermodynamic potential of the system
at high temperature
by neglecting the zero-point energy
\begin{equation}\label{33}
\Omega(\beta,A)=-\mathrm{L}(A)+F(\beta,A)=
\frac 14 \int_{{\cal M}^3} d^3x F_{\mu\nu}F^{\mu\nu}+F(\beta,A)~~,
\end{equation}
where $\mathrm{L}(A)$ is the classical Lagrangian of the background
potential $A_\mu$.
Eqs. (\ref{28}), (\ref{30})--(\ref{32}) yield
\begin{equation}\label{35}
\Omega(\beta,A)= -\int_{{\cal M}^3} d^3x \left(-{c(T) \over 4} F_{\mu\nu}F^{\mu\nu}
+\frac 12 m^2(T)A_0^2\right)~,
\end{equation}
\begin{equation}\label{36}
c(T)=1-{e^2 \over 24\pi^2}\ln (T/\rho)~~,~~
m^2(T)=\frac 13 e^2T^2~.
\end{equation}
These formulas reproduce the result which is well-known in plasma physics.
The last term in (\ref{35}) is responsible for
screening effects in hot plasma and $m(T)$
is the correct expression for the Debye mass.

Note that conventional derivation of (\ref{35}) is based on computation
of one-loop
diagrams at finite temperature in the limit of vanishing gauge fields.
Our calculation is done for arbitrary static gauge potential.

Coefficient $a_2$ in (\ref{32}) has a Lorentz invariant structure.
For this reason the logarithmic term in expansion (\ref{28}) results in a
finite renormalization of the background potential (see the first term in r.h.s.
of (\ref{35})). The Lorentz invariant structure of $a_2$ is not a mere coincidence.
It reflects a more general property established in \cite{df:02a}:
coefficient $a_2$ of the pseudo-trace expansion coincides with the coefficient
$A_2$ of the asymptotic expansion of the heat kernel of the covariant
four-dimensional operator $P(i\partial_0)
=-\partial_0^2-L(i\partial_0)$.

\section{Operator Polynomials for Different Field Theories}
\setcounter{equation}0

Spectral asymptotics given in Sec. 2 are obtained by using certain
conditions. In particular, the structure of
coefficients $a_n$ and $b_n$ in (\ref{20}), (\ref{21}) is determined by
condition (iii).
Let us analyze the origin of this condition for models of non-interacting
fields.

Consider a pair $f_1$, $f_2$ of solutions to
field Eq. (\ref{2}) (or (\ref{4}), for Dirac case)
and introduce the relativistic product, a Hermitian bilinear form
$\langle f_1, f_2 \rangle=
\langle f_2, f_1 \rangle^{*}$
defined on a space-like hypersurface
$\Sigma$ in a hyperbolic spacetime $\cal M$ as
\begin{equation}\label{37}
\langle f_1, f_2 \rangle=\int_{\Sigma}
d\Sigma^\mu~j_\mu (f_1, f_2)~.
\end{equation}
The "current" $j_\mu (f_1, f_2)$ is divergence free,
$\nabla^\mu j_\mu=0$, and (\ref{37}) does not depend on $\Sigma$.

If the background is  stationary
one can introduce a conserved canonical energy $H[f]$ for a given solution
$f$.
It can be shown that for free scalar, spinor and gauge fields
the energy can be written in the form
\begin{equation}\label{38}
H[f]=\frac i2 \langle f, \partial_0 f \rangle+c.c.~,
\end{equation}
where "$c.c.$"
is a term obtained from the first term in the r.h.s. of (\ref{38})
by complex conjugation. It follows from (\ref{38}) that
for a solution $f_\omega$ with certain frequency $\omega$
($i\partial_0 f_\omega=\omega f_\omega$)
\begin{equation}\label{39}
H[f_\omega]=\omega \langle f_\omega, f_\omega \rangle~.
\end{equation}
In general the system may have excitations with complex frequencies $\omega$.
It is easy to see that they have vanishing relativistic norm
$\langle f_\omega, f_\omega \rangle$ and, hence,
the vanishing energy. Such modes along with zero-frequency
modes are not quantized and should be considered separately from real-frequency
solutions.  This is the reason  why pseudo-trace (\ref{16}) is defined
for real $\omega$.

\subsection{Scalar Fields}

Consider integer spin  fields.
To have a well-defined theory we require that canonical energy is non-negative,
\begin{equation}\label{40}
H[f_\omega]\geq 0~~,
\end{equation}
for any $\omega$. If there are single-particle
excitations with negative energy $H$ the Fock space in the
corresponding quantized theory contains states with negative energies unbounded
from below.

Condition (iii) of Sec. 2 for scalar fields $\phi$ follows from (\ref{40}).
Indeed it is possible to show \cite{df:02a} that in this case
\begin{equation}\label{41}
\langle \phi_\omega, \phi_\omega \rangle=\chi'(\omega)(\phi_\omega, \phi_\omega)~,
\end{equation}
where $\chi'(\omega)$ was introduced in (\ref{11}) and
$(\phi_\omega, \phi_\omega)=\int dV |\phi_\omega|^2$ is a positive-definite norm in a Hilbert space
of single-particle wave functions.
(\ref{40}) together with relations (\ref{39}), (\ref{41})
implies that $\chi'(\omega)=\varepsilon(\omega) |\chi'(\omega)|$ where
$\varepsilon(\omega)$ is the sign function.

\subsection{Gauge Fields}

We discuss now theory of non-Abelian gauge fields with
the gauge group $SU(N)$.
Let $B_\mu$ be a solution to Yang-Mills equations, ${\cal D}^\mu F^{(B)}_{\mu\nu}=0$,
where $F^{(B)}_{\mu\nu}$ is the strength tensor of
$B_\mu$ (we work in the adjoint representation).
In the linear order, perturbations $A_\mu$ of the Yang-Mills
field on the background $B_\mu$ obey the equation
\begin{equation}\label{42}
{\cal D}^\nu{\cal D}_\nu A_\mu-{\cal D}_\mu{\cal D}^\nu A_\nu+2iF^{(B)}_{\mu\nu}A^\nu=0~,
\end{equation}
where ${\cal D}_\mu=\partial_\mu+iB_\mu$.
It is convenient to work in the background gauge ${\cal D}_\mu A^\mu=0$
where equation transforms to
\begin{equation}\label{42a}
{\cal D}^\nu{\cal D}_\nu A_\mu+2iF^{(B)}_{\mu\nu}A^\nu=0~~.
\end{equation}
Fixing this gauge leaves a freedom in the gauge transformations
$\delta A_\mu={\cal D}_\mu \lambda$ where $\lambda$ is a solution to
\begin{equation}\label{43}
{\cal D}_\mu {\cal D}^\mu\lambda=0~~.
\end{equation}
Suppose now that $B_\mu$ is a stationary field. Then
Eqs. (\ref{42})--(\ref{43}) result in operator polynomials
of the form (\ref{3}).
Let $\omega$ be the physical spectrum of (\ref{42}),
$\omega_{(1)}$ the spectrum
related to vector Eq. (\ref{42a}) and $\omega_{(0)}$ the spectrum
of (\ref{43}).
It is clear that $\omega$'s represent a subset among eigenvalues $\omega_{(1)}$.
The pseudo-trace for the physical spectrum
can be represented as
\begin{equation}\label{44}
K(t)=\sum_{\omega>0}e^{-t\omega^2}=K_{(1)}(t)-2K_{(0)}(t)
\equiv
\sum_{\omega_{(1)}>0}e^{-t\omega_{(1)}^2}-2\sum_{\omega_{(0)}>0}e^{-t\omega_{(0)}^2}~~.
\end{equation}
(For $SU(N)$ group Eqs. (\ref{42})--(\ref{43})
are invariant
with respect to the charge conjugation and their spectrum is
symmetric with respect to change $\omega$ to $-\omega$.)
The double subtraction of the sum over $\omega_{(0)}$ in (\ref{44}) has the
following explanation.
One subtraction eliminates
solutions of (\ref{42a}) which do not respect the gauge
conditions, ${\cal D}_\mu A^\mu=\phi\neq 0$. It is easy to see that
such modes have frequencies $\omega_{(0)}$ because
$\phi$ obeys scalar Eq. (\ref{43}).
Additional subtraction of $K_{(0)}(t)$ in (\ref{44}) eliminates pure gauge solutions
$A_\mu={\cal D}_\mu \lambda$.
The last term in r.h.s of (\ref{44})
can be interpreted as contribution of ghosts.

There is an analog of formula (\ref{41}) for transverse
solutions (${\cal D}_\mu A^\mu=0$) to (\ref{42}).  The relativistic norm
for the solutions with energy $\omega$ can be written as
\begin{equation}\label{47}
\langle A_\omega,A_\omega \rangle=
\chi'(\omega)(A_\omega,A_\omega)~~,
\end{equation}
where $(A_\omega,A_\omega)=\int dV (A_\omega^{+})^\mu (A_\omega)_\mu$
and quantity $\chi'(\omega)$ is defined for the polynomial
corresponding to vector problem (\ref{42a}).
The residual gauge freedom can be used to impose condition $A_0=0$.
For such solutions the norm $(A_\omega,A_\omega)$ becomes positive-definite.

Suppose that the canonical energy of perturbations $A_\mu$
on the background $B_\mu$ is non-negative,
$H_{YM}[A]\geq 0$. Require also the same property
for canonical energy of the scalar field
described by (\ref{43}), $H[\lambda]\geq 0$.
Then one can show that
operator polynomials associated with vector, (\ref{42a}), and scalar,
(\ref{43}), problems obey condition (iii) of Sec. 2. The proof is based on relations
(\ref{41}), (\ref{47}).

\subsection{Dirac Field}

As is known, the classical energy for the Dirac field is not positive-definite.
Thus, the proof of condition (iii) in this case should be different.
Consider the Dirac field in Minkowsky space-time in the presence
of a static gauge potential $A_\mu$.
The relativistic norm of a single-particle wave function is
$\langle\psi,\psi\rangle=\int dV~\psi^{+}\psi$.
If the field has a mass, $m\neq 0$, one can prove the relation \cite{df:02a}
\begin{equation}\label{45}
\langle \psi_\omega,\psi_\omega\rangle= {\chi'(\omega_1) \over 2m}
(\psi_\omega,\psi_\omega)~,
\end{equation}
where $(\psi_\omega,\psi_\omega)=\int dV ~\psi_\omega^{+}i\gamma_0\psi_\omega$
and $\gamma_0$ is anti-Hermitian.
It follows from (\ref{45}) and  positivity of the norm $\langle\psi,\psi\rangle$
that  $\chi'(\omega)$ does not vanish for massive fields.
Now, if $A_\mu=0$, one has $\chi'(\omega)=2\omega$ and condition (iii) is satisfied.
The validity of (iii) for $A_\mu\neq 0$ can be proved by continuity.
Because  $\chi'(\omega)\neq 0$ quantity $\chi'(\omega)$ cannot change the sign when
$A_\mu$ is switched on from zero to some value.

\section*{Acknowledgments}

The author is grateful to the Organizing Committee of the 6th Workshop
on "Quantum Field Theory under the Influence of External Conditions"
for the
financial support.

%\newpage

\end{document}